\documentclass{elsart}
\usepackage{graphics,harvard,amssymb}
\journal{New Astronomy}

\makeatletter
\def\elsartstyle{%
	\def\normalsize{\@setfontsize\normalsize\@xiipt{14.5}}
	\def\small{\@setfontsize\small\@xipt{13.6}}
	\let\footnotesize=\small
	\def\large{\@setfontsize\large\@xivpt{18}}
	\def\Large{\@setfontsize\Large\@xviipt{22}}
	\skip\@mpfootins = 18\p@ \@plus 2\p@
	\normalsize
}
\makeatother

\def\bibcode#1{(\texttt{#1})}
\def\astrobj#1{#1}
\def\url#1{{\ttfamily\def\/{/\discretionary{}{}{}}#1}}

\begin{document}

\begin{frontmatter}
\title{Models of \astrobj{Procyon A} including seismic constraints}
      
\author[Geneve]{P. Eggenberger\thanksref{email}},
\author[Geneve]{F. Carrier},
\author[Geneve,Marseille]{F. Bouchy}
\address[Geneve]{Observatoire de Gen\`eve, 51 Ch. des Maillettes, CH--1290 Sauverny, Suisse}
\address[Marseille]{Laboratoire d'Astrophysique de Marseille, Traverse du Siphon, BP 8, 13376 Marseille Cedex 12, France}

\thanks[email]{E-mail: Patrick.Eggenberger@obs.unige.ch}

\begin{abstract}
Detailed models of \astrobj{Procyon~A} based on new asteroseismic measurements by \citeasnoun{eg04b} have been computed 
using the Geneva evolution code including shellular rotation and atomic diffusion. By combining all
non--asteroseismic observables now available for \astrobj{Procyon~A} with these seismological data, we find 
that the observed mean large spacing of $55.5 \pm 0.5$\,$\mu$Hz 
favours a mass of 1.497\,$M_{\odot}$ for \astrobj{Procyon~A}. We also determine the following global parameters of \astrobj{\astrobj{Procyon~A}}:
an age of $t=1.72 \pm 0.30$\,Gyr, an initial helium mass fraction $Y_{\mathrm{i}}=0.290 \pm 0.010$, 
a nearly solar initial metallicity $(Z/X)_{\mathrm{i}}=0.0234 \pm 0.0015$
and a mixing--length parameter $\alpha=1.75 \pm 0.40$.
Moreover, we show that the effects of rotation on the inner structure of the star may be revealed by asteroseismic observations
if frequencies can be determined with a high precision. Existing seismological data of \astrobj{\astrobj{Procyon~A}} are unfortunately 
not accurate enough to really test these differences in the input physics of our models.
\end{abstract}

\end{frontmatter}

\section{Introduction}

Due to its brightness and proximity, \astrobj{Procyon~A} constitutes an ideal target to
test the input physics of the stellar models and to 
search for $p$--mode oscillations.

\citeasnoun{ha75} were the first to calculate stellar models of \astrobj{Procyon~A} and to compare them to the observational data available 
at the time. They found a discrepancy between the astrometric mass \cite{st51} and the 
lower astrophysical mass deduced from their models. This mass discrepancy was later confirmed by
\citeasnoun{de88}, who also found that overshoot of the convective core had to be included in the models
in order to reproduce the observed effective temperature. 

To adress this problem of mass discrepancy, \citeasnoun{ir92}
decided to redetermine the astrometric mass of \astrobj{Procyon~A}.
They found a mass in perfect agreement with the previous value of 1.74\,$M_{\odot}$.
Following this redetermination of Procyon's mass and motivated by several attempts made to
detect the signature of oscillation modes, \citeasnoun{gu93} calculated an array of stellar models for
\astrobj{Procyon~A} including the new OPAL opacities. They concluded that, using OPAL opacities, no convective overshoot
was needed to match Procyon's position in the HR diagram. However, the discrepancy between the astrometric mass
and the mass deduced from stellar evolution models still remained.

This discrepancy was solved thanks to
new measurements of orbital elements and parallax of the Procyon system by Girard et al. (1996, 2000).
They determined an astrometric mass of 1.5\,$M_{\odot}$ in good agreement with the value supported by
stellar models. Using this revised astrometric mass, \citeasnoun{ch99} calculated a grid of stellar evolution models
for \astrobj{Procyon~A} and investigated their seismic properties. They concluded that the detection of $p$--modes
would serve as a robust test of stellar evolution theory. 

The first indication of the presence of $p$--modes on \astrobj{Procyon~A} was obtained by \citeasnoun{bo91}, while
the first clear detection was made by \citeasnoun{ma99}. 
By comparing these observations with
theoretical predictions and numerical simulations, \citeasnoun{ba99} confirmed the stellar origin of the observed
excess power.

Very recently, individual $p$--mode frequencies were identified (Eggenberger et al. 2004b, hereafter ECBB04;
Marti\'c et al. 2004). 
However, \citeasnoun{mat04} reported that the \textsc{Most} satellite did not
observed any evidence of the expected acoustic oscillations. To explain the apparent discrepancy with previous 
radial--velocity
measurements, Matthews et al. suggested that either both the radial velocity measurements and the \textsc{Most} observations
are dominated by granulation noise, or the properties of the oscillations are different from the theoretical
expectations. \citeasnoun{ch04} cast doubt on the fact that granulation dominates the noise, and 
suspected that the \textsc{Most} data might be dominated by non--stellar noise.
Moreover, new asteroseismic observations with the \textsc{Harps} spectrograph confirmed the previous
Doppler ground-based detections \cite{bo04}.

In this work, we will combine all non--asteroseismic measurements with asteroseismic observations of ECBB04
to investigate which additional constraints are brought by these seismological data. We will thus try to
determine a model of \astrobj{Procyon~A} which best reproduces all these observational constraints using the Geneva
evolution code which includes a complete treatment of shellular rotation and atomic diffusion. 
Moreover, we will investigate the effects of rotation on the global parameters of \astrobj{Procyon~A} and on the 
$p$--mode frequencies.

The observational constraints available for \astrobj{Procyon~A} are summarized in Sect.~2, while the input physics of the models
and the calibration method are described in Sect.~3. The results are presented in Sect.~4 and the conclusion is given 
in Sect.~5.

\section{Observational constraints}

\label{obs}

\subsection{Astrometric data}
\label{astro}

The astrometric parameters of the visual binary orbit were recently updated by \citeasnoun{gi00}.
Using data obtained with the infrared cold coronagraph (CoCo) and with the WFPC2 on board the Hubble Space Telescope,
they found a mass of $1.497 \pm 0.037$\,$M_{\odot}$ for \astrobj{Procyon A}, with a parallax $\Pi=283.2 \pm 1.5$\,mas.
When only the WFPC2 observations are used, a smaller mass is found: $M=1.465 \pm 0.041$\,$M_{\odot}$.
These masses were derived by using the self--consistent parallax of $283.2 \pm 1.5$\,mas.
Using the slightly larger parallax measured by Hipparcos ($\Pi=285.93 \pm 0.88$\,mas) 
with the astrometric parameters of the orbit results in a smaller mass than the one obtained
with the parallax of \citeasnoun{gi00}.
Indeed, when only the WFPC2 measurements are used, we find a mass of $1.423 \pm 0.040$\,$M_{\odot}$ with the
Hipparcos parallax. Thus, we see that the mass of \astrobj{Procyon A} lies between 1.42 and 1.5\,$M_{\odot}$.
All of these masses will be considered in our analysis. We will then determine which value is in best accordance with the other observational
constraints and in particular with the asteroseismic measurements.

\subsection{Effective temperature and chemical composition}
\label{tchim}

For the effective temperature of \astrobj{Procyon~A}, we adopted 
$T_{\mathrm{eff}}=6530 \pm 90$\,K \cite{fu97}.
Note that the value of 6530\,K is also given by \citeasnoun{al02}.
The chemical composition of \astrobj{Procyon~A} is nearly solar;
we adopted 
$[\mathrm{Fe/H}]=-0.05 \pm 0.03$ \cite{al02}.

\subsection{Luminosity}
\label{lum}

From the compilation of 13 measurements from the literature, \citeasnoun{al02}
derived a mean visual magnitude $\langle V \rangle = 0.363 \pm 0.003$\,mag for \astrobj{Procyon~A}.
Combining this mean magnitude  
with the solar absolute bolometric magnitude $M_{\mathrm{bol},\,\odot}=4.746$
\cite{le98} and the
bolometric correction from
\citeasnoun{flower}, 
we find a luminosity $\log L/L_{\odot} = 0.84 \pm 0.02$.
Note that this interval in luminosity is compatible with the use of the parallax of 
\citeasnoun{gi00} as well as the Hipparcos parallax. Indeed, the changes in $L$
due to the different parallaxes are very small and entirely included in the error bar of
0.02 ($\log L/L_{\odot} = 0.843$ and $0.835$ with the parallax of Girard et al. and the 
Hipparcos parallax respectively).

\subsection{Angular diameter}

Recently \citeasnoun{ke04} measured the angular diameter of \astrobj{Procyon~A} using the VINCI instrument
installed at ESO's VLT Interferometer. They found a limb darkened angular diameter 
$\theta=5.448 \pm 0.053$\,mas.
Using the Hipparcos parallax, they deduced a linear diameter of $2.048 \pm 0.025$\,$D_{\odot}$.
The parallax of \citeasnoun{gi00} gives a linear diameter of $2.067 \pm 0.028$\,$D_{\odot}$.

\subsection{Rotational velocity}

\citeasnoun{al02} estimated $v \sin i =3.16 \pm 0.50$\,km\,s$^{-1}$. However, they
pointed out that the correct value is probably close to $2.7$\,km\,s$^{-1}$,
since the value of 3.16 may be slightly overestimated as a result of the finite
numerical resolution of their convection simulation.
Thus, we adopted $v \sin i=2.7$\,km\,s$^{-1}$ with a large error of 1\,km\,s$^{-1}$ in order
to encompass the value of $v \sin i =3.16 \pm 0.50$\,km\,s$^{-1}$.
Using $v \sin i=2.7 \pm 1.0$\,km\,s$^{-1}$ and assuming that the rotation axis of \astrobj{Procyon~A}
is perpendicular to the plane of the visual orbit ($i=31.1\pm 0.6^{\circ}$, Girard et al. 2000), 
we find a surface rotational
velocity of $5.2 \pm 1.9$\,km\,s$^{-1}$ for \astrobj{Procyon~A}.

\begin{table}
\caption[]{Observational constraints for \astrobj{Procyon~A}. References: (1) \citeasnoun{gi00}, 
(2) Hipparcos,
(3) \citeasnoun{al02},
(4) derived from the other observational measurements (see text), 
(5) \citeasnoun{fu97},
(6) \citeasnoun{ke04},
(7) \citeasnoun{ma04} and
(8) \citeasnoun{eg04b}.}

\begin{center}
\begin{tabular}{ccc}
\hline
\hline
 &  & References \\ \hline
$\Pi$ [mas]& $283.2 \pm 1.5$   & (1) \\
           & $285.93 \pm 0.88$ & (2) \\
$M/M_{\odot}$  & $1.497 \pm 0.037$ & (1) \\
               & $1.465 \pm 0.041$ & (1) \\
               & $1.423 \pm 0.040$ & (1)+(2) \\
$V$ [mag] & $0.363 \pm 0.003$ & (3) \\
$\log L/L_{\odot}$ & $0.84 \pm 0.02$ & (4) \\
$T_{\mathrm{eff}}$ [K]& $6530 \pm 90$ & (5) \\
$[$Fe/H$]_{\mathrm{s}}$ &  $-0.05 \pm 0.03$ & (3) \\
$\theta$ [mas] &  $5.448 \pm 0.053$ & (6)\\
$R/R_{\odot}$ & $2.048 \pm 0.025$ & (6)+(2) \\
              & $2.067 \pm 0.028$ & (6)+(1) \\
$V_{\mathrm{s}}$ [km\,s$^{-1}$] & $5.2 \pm 1.9$ & (4) \\
$\Delta \nu_0$ [$\mu$Hz] & $53.6 \pm 0.5$ & (7) \\
                         & $55.5 \pm 0.5$ & (8) \\
 
\hline
\label{tab:constraints}
\end{tabular}
\end{center}
\end{table}

\subsection{Asteroseismic constraints}
\label{asc}

\citeasnoun{ma04} identified individual frequencies in the 
power spectrum between 300 and 1400\,$\mu$Hz with a mean large spacing of $53.6 \pm 0.5$\,$\mu$Hz.
Using the \textsc{Coralie} spectrograph, we confirmed the detection of p--modes on \astrobj{Procyon~A}
and identified individual frequencies with a slightly larger 
mean large spacing of $55.5 \pm 0.5$\,$\mu$Hz (ECBB04).
In this work, we will use the asteroseismic observations of ECBB04 to constrain the models.
  
All the observational constraints are listed in Table~\ref{tab:constraints}.

\section{Stellar models}

\subsection{Input Physics}

The stellar evolution code used for these computations is the Geneva code
including shellular rotation, described several times in the literature 
\citeaffixed{mm00}{see}. 
We used the new horizontal turbulence prescription of \citeasnoun{ma03} and the braking law
of \citeasnoun{ka88} in order to reproduce the magnetic braking that undergo the low
mass stars when arriving on the main sequence. Two parameters enter this braking law: the
saturation velocity $\Omega_{\rm{sat}}$ and the braking constant $K$. Following \citeasnoun{bo97},  
$\Omega_{\rm{sat}}$ was fixed to 14\,$\Omega_{\odot}$ and the braking constant $K$
was calibrated on the sun.
We used the OPAL opacities, 
the NACRE nuclear reaction
rates \cite{an99} and the standard mixing--length formalism for convection. 

In addition to shellular rotation, our models have been computed including atomic diffusion on
He, C, N, O, Ne and Mg using the
routines developed for the Geneva--Toulouse version of our code
\citeaffixed{ri96}{see for example} recently updated by O. Richard
(private communication).
The diffusion coefficients are computed with the prescription by
\citeasnoun{pa86}.
We included the diffusion due to the concentration and thermal
gradients, but the radiative acceleration was neglected.

\begin{figure}[htb!]
 \resizebox{\hsize}{!}{\includegraphics{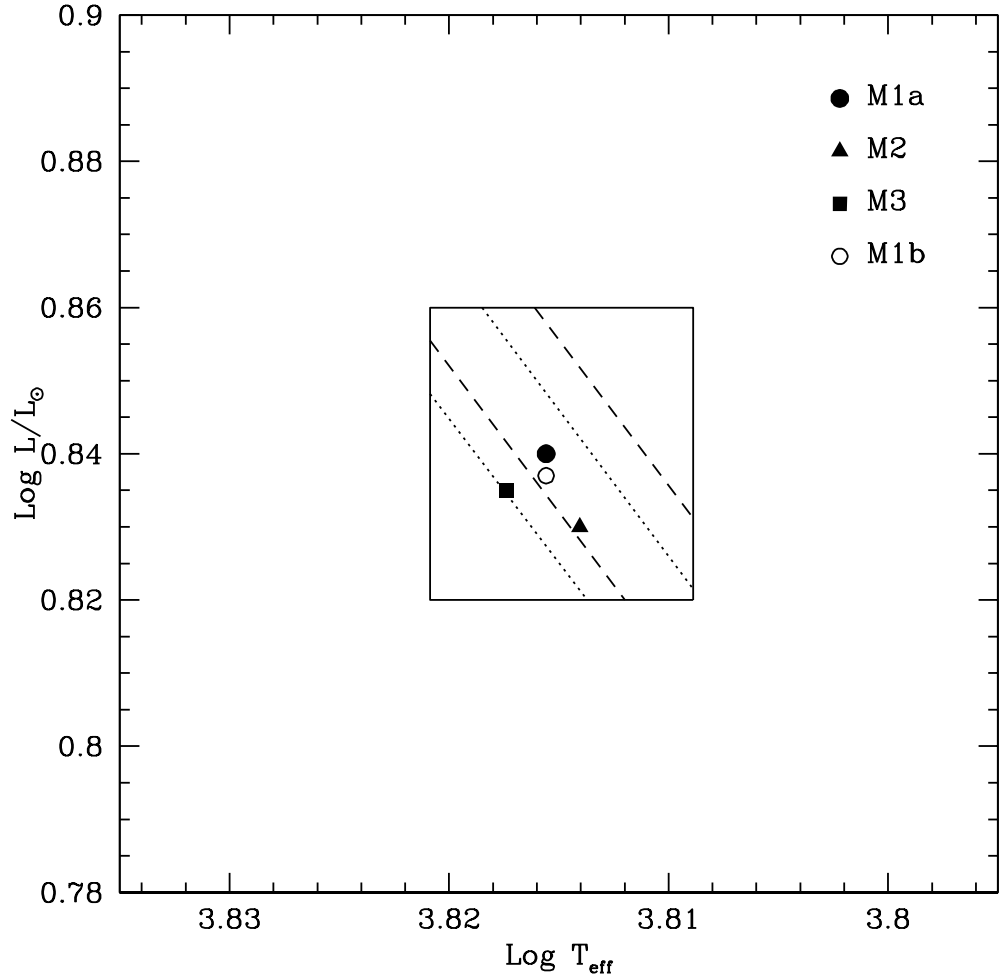}}
  \caption[]{Observational constraints in the HR diagram for \astrobj{Procyon~A}. The continuous lines indicate the boxes delimited
  by the observed luminosity and effective temperature (with their respective 1--sigma errors). The dashed and dotted lines
  denote the boxes delimited by the interferometric radii deduced using the parallax of \citeasnoun{gi00} and the Hipparcos
  parallax, respectively.
  The positions in the HR diagram of the computed models are indicated (see text and Table \ref{tab:res} for more details).}
  \label{dhr}
\end{figure}

\subsection{Computational method}

The computation of a stellar model for a given star consists in finding the set of stellar modeling parameters which best reproduces all observational 
data available for this star.
For a given stellar mass, the characteristics of a stellar model including the effects of rotation (luminosity, effective temperature, surface metallicity, surface velocity, frequencies
of oscillation modes, etc.) depend on five modeling parameters: the age of the star ($t$ hereafter), 
the mixing--length parameter $\alpha \equiv l/H_{\mathrm{p}}$ for convection, the initial surface velocity $V_{\mathrm{i}}$
and two parameters describing the initial chemical composition of the star. For these two parameters,
we chose the initial helium abundance $Y_{\mathrm{i}}$ and the initial ratio between the mass fraction of heavy elements and hydrogen 
$(Z/X)_{\mathrm{i}}$. 
Assuming that this ratio is proportional to the abundance ratio [Fe/H], we can directly relate $(Z/X)$ to [Fe/H]
by using the solar value $(Z/X)_{\odot}=0.0230$ given by \citeasnoun{gr98}.
Thus, any characteristic $A$ of a given stellar model has the following formal dependences with respect to modeling parameters :
$A=A(t,\alpha,V_{\mathrm{i}},Y_{\mathrm{i}},(Z/X)_{\mathrm{i}})$.\\

Once fixing the mass of \astrobj{Procyon~A} to one of the three values listed in Table~\ref{tab:constraints}, the determination of the set of 
modeling parameters ($t$, $\alpha$, $V_{\mathrm{i}}$, $Y_{\mathrm{i}}$, $(Z/X)_{\mathrm{i}}$) leading to the best agreement with the observational constraints
is made in two steps. 
First, we construct a grid of models with position in the HR diagram
in agreement with the observational values of the luminosity, effective temperature and radius listed in Table~\ref{tab:constraints}.
The boxes in the HR diagram for \astrobj{Procyon~A} corresponding to the three different choices of mass are shown in Fig.~\ref{dhr}. 

To construct this grid, we proceed in the following way: for a given chemical composition
(i.e. a given set $Y_{\mathrm{i}}$, $(Z/X)_{\mathrm{i}}$) and a given initial velocity $V_{\mathrm{i}}$, the mixing--length coefficient of each star is adjusted in order to match the observational
position in the HR diagram. 
     
Once the position of the star in the HR diagram agrees with the observed values, the surface metallicity of
the star is compared to the observed one. If it is out of the metallicity interval listed in 
Table~\ref{tab:constraints}, the models are rejected and the procedure is repeated with another choice of $Y_{\mathrm{i}}$ and $(Z/X)_{\mathrm{i}}$.
Note that the surface metallicities [Fe/H]$_{\mathrm{s}}$ are almost identical for the models 
with the same initial composition and different mixing--length parameters.
Moreover, the [Fe/H]$_{\mathrm{s}}$ of the models are mainly sensitive to $(Z/X)_{\mathrm{i}}$ and less to $Y_{\mathrm{i}}$. 
As a result,
the values of $(Z/X)_{\mathrm{i}}$ are directly constrained by the observed surface metallicity.

Finally, the surface velocity of the models is compared to the observed one; if the velocities of the models are not compatible with the observed value,
the models are rejected and the procedure is repeated with another initial velocity (but with the same initial chemical composition).
Otherwise, all solutions are kept, since they correspond to models
of \astrobj{Procyon A} which reproduce all the non--asteroseismic constraints. The whole procedure is then 
repeated with a new choice 
of $Y_{\mathrm{i}}$ and $(Z/X)_{\mathrm{i}}$.

In this way we obtain a grid of models with various sets of modeling parameters ($t$, $\alpha$, $V_{\mathrm{i}}$, $Y_{\mathrm{i}}$, $(Z/X)_{\mathrm{i}}$)
which satisfy all the non--asteroseismic observational constraints of \astrobj{Procyon~A}, namely the effective temperature,
the luminosity, the radius, the surface velocity and the surface metallicity. 
The second step in determining the best model of \astrobj{Procyon A} is to consider
the asteroseismic measurements. 
    
For each stellar model of the grid constructed as explained above, low--$l$ p--mode frequencies are calculated using the Aarhus adiabatic pulsation
package \cite{cd97}. Following our observations, modes of degree $l \leq 2$ with frequencies between 0.6 and 1.4 mHz
are computed and the large spacing $\Delta \nu_0$ determined.
  
Once the asteroseismic characteristics of all relevant models are computed, we perform a $\chi^2$ minimization in order to deduce the 
set of parameters ($t$, $\alpha$, $V_{\mathrm{i}}$, $Y_{\mathrm{i}}$, $(Z/X)_{\mathrm{i}}$) leading to the 
best agreement with the observations of \astrobj{Procyon A}. For this purpose, we define the $\chi^2$ functional
\begin{eqnarray}
\label{eq1}
\chi^2 \equiv \sum_{i=1}^{6} \left( \frac{C_i^{\mathrm{theo}}-C_i^{\mathrm{obs}}}{\sigma C_i^{\mathrm{obs}}} \right)^2  \; ,
\end{eqnarray}
where the vectors $\mathbf{C}$ contains all the observables for the star:  
\begin{eqnarray}
\nonumber
\mathbf{C} \equiv (\log L/L_{\odot},T_{\mathrm{eff}},R/R_{\odot},V_{\mathrm{s}},
[\mathrm{Fe/H}]_{\mathrm{s}},\Delta \nu_0) \; .   
\end{eqnarray} 
The vector $\mathbf{C}^{\mathrm{theo}}$ contains the theoretical values of these observables for the model to be tested, while 
the values of $\mathbf{C}^{\mathrm{obs}}$ are those
listed in Table~\ref{tab:constraints}. The vector $\mathbf{\sigma C}$ contains the errors on these observations which are also given in
Table~\ref{tab:constraints}.

\section{Results}

\subsection{Models with a mass of $1.497\,M_{\odot}$}

We first calculated a grid of models including shellular rotation and atomic diffusion 
with the mass of $1.497\,M_{\odot}$ deduced from data obtained with
the infrared cold coronagraph (CoCo) and with the WFPC2 on board the Hubble Space Telescope \citeasnoun{gi00}.
This mass has been determined with a parallax of $283.2 \pm 1.5$\,mas, leading to
a radius of $2.067 \pm 0.028$\,$R_{\odot}$ for \astrobj{Procyon~A}.
The models have to match the location of \astrobj{Procyon~A} in the HR diagram
which is given by its radius, luminosity and effective temperature (see Fig.~\ref{dhr}). 

\begin{figure}[htb!]
 \resizebox{\hsize}{!}{\includegraphics{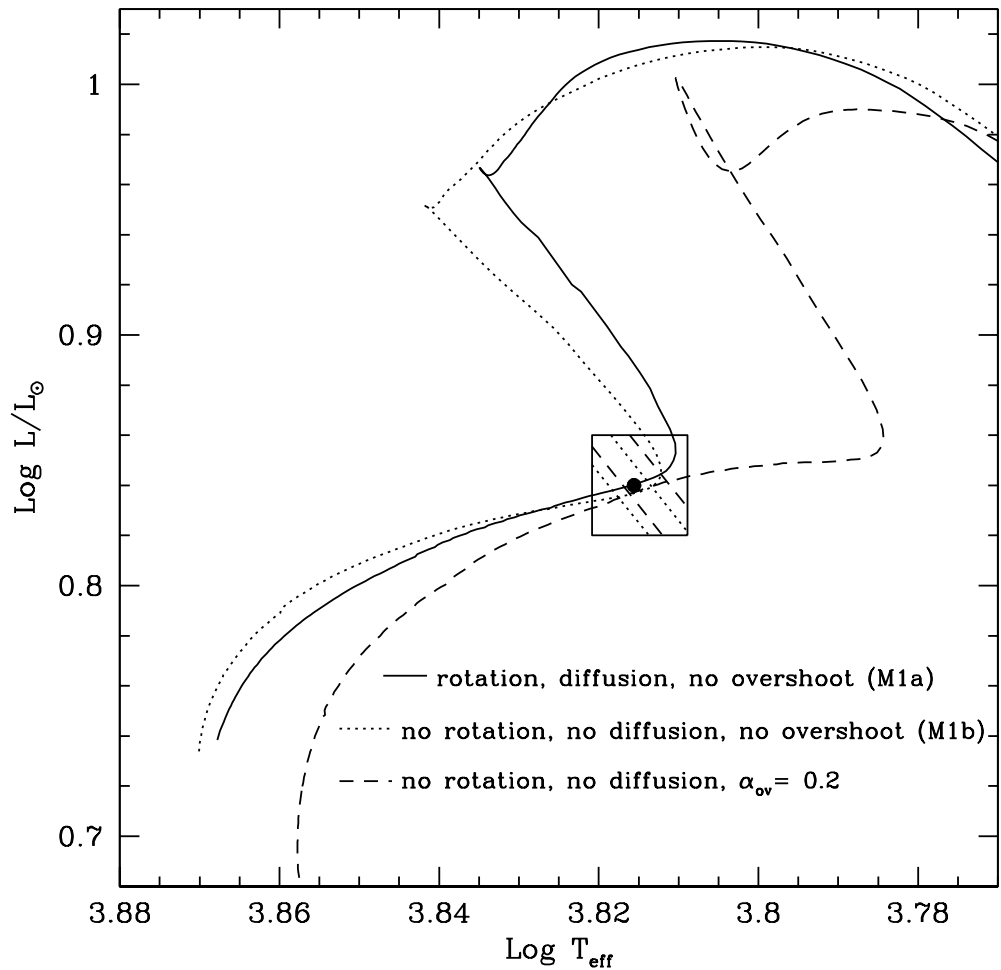}}
  \caption[]{Evolutionary tracks in the HR diagram for three models of 1.497~$M_{\odot}$.
   The continuous line corresponds to the model including rotation and atomic diffusion but no overshooting
  (model M1a). The dotted line denotes the model without rotation, atomic diffusion nor overshooting (model M1b), while 
  the dashed line corresponds to a model computed with overshooting ($\alpha_{\mathrm{ov}}=0.2$), but without rotation and atomic diffusion. 
  The dot shows the location of the M1a model.}
  \label{dhr_ov}
\end{figure}

Once this grid of models was computed, we performed the $\chi^2$ minimization described above.
In this way, we found the solution
$t=1.72 \pm 0.30$\,Gyr, $\alpha=1.75 \pm 0.40$, $V_{\mathrm{i}}=14 \pm 8$\,km\,s$^{-1}$, 
$Y_{\mathrm{i}}=0.290 \pm 0.010$ and $(Z/X)_{\mathrm{i}}=0.0234 \pm 0.0015$. The position and the evolutionary track of this model 
(denoted model M1a in the following) in the HR diagram are shown in Fig.~\ref{dhr} and \ref{dhr_ov}. 
The characteristics of this model
are reported in Table~\ref{tab:res}. 
Note that the confidence limits of each modeling parameter given in Table~\ref{tab:res} are estimated as the maximum/minimum values which 
fit the observational constraints when the other calibration parameters are fixed to their medium value.  

\begin{table*}
\caption[]{Models for \astrobj{Procyon A}. The upper part of the table gives the observational constraints used for the
calibration. The middle part of the table presents the modeling parameters with their confidence limits, while the bottom
part presents the global parameters of the star.}
\begin{center}
\label{tab:res}
\begin{tabular}{c|ccc|c}
\hline
\hline
 & \multicolumn{3}{c}{Models including rotation} & \multicolumn{1}{|c}{Model without rotation}  \\
 & \multicolumn{3}{c}{and atomic diffusion} & \multicolumn{1}{|c}{and diffusion} \\
 & \multicolumn{1}{c}{M1a} & \multicolumn{1}{c}{M2} & \multicolumn{1}{c|}{M3} & \multicolumn{1}{c}{M1b} \\ \hline
$M/M_{\odot}$  & $1.497$ & $1.465$ & $1.423$ & $1.497$ \\
$\log L/L_{\odot}$ & $0.84 \pm 0.02$ & $0.84 \pm 0.02$ & $0.84 \pm 0.02$ & $0.84 \pm 0.02$ \\
$T_{\mathrm{eff}}$ [K]& $6530 \pm 90$ & $6530 \pm 90$ & $6530 \pm 90$ & $6530 \pm 90$ \\
$R/R_{\odot}$ & $2.067 \pm 0.028$ & $2.067 \pm 0.028$ & $2.048 \pm 0.025$ & $2.067 \pm 0.028$ \\
$[$Fe/H$]_{\mathrm{s}}$ & $-0.05 \pm 0.03$ & $-0.05 \pm 0.03$ &  $-0.05 \pm 0.03$ & $-0.05 \pm 0.03$ \\
$V_{\mathrm{s}}$ [km\,s$^{-1}$] & $5.2 \pm 1.9$ & $5.2 \pm 1.9$ & $5.2 \pm 1.9$ & $ - $\\
$\Delta \nu_{0}$ [$\mu$Hz] & $55.5 \pm 0.5$  & $55.5 \pm 0.5$ & $55.5 \pm 0.5$  & $55.5 \pm 0.5$\\
\hline
$t$ [Gyr] &  $1.72 \pm 0.30$ & $1.89 \pm 0.30$ & $2.18 \pm 0.30$ & $1.77 \pm 0.30$\\
$\alpha$ & $1.75 \pm 0.40$ & $1.90 \pm 0.30$ & $1.80 \pm 0.30$ & $1.60 \pm 0.35$  \\
$V_{\mathrm{i}}$ [km\,s$^{-1}$] & $14 \pm 8$ & $17 \pm 8$ & $17 \pm 8$ & $ - $\\
$Y_{\mathrm{i}}$  & $0.290 \pm 0.010$ & $0.295 \pm 0.015$ & $0.295 \pm 0.020$ & $0.280 \pm 0.010$\\
$(Z/X)_{\mathrm{i}}$ & $0.0234 \pm 0.0015$ & $0.0229 \pm 0.0015$ & $0.0231 \pm 0.0015$ & $0.0205 \pm 0.0015$  \\
\hline
$\log L/L_{\odot}$ & 0.840 & 0.830 & 0.835 & 0.837 \\
$T_{\mathrm{eff}}$ [K]& $6540$ & $6517$ & $6567$ & $6540$ \\
$R/R_{\odot}$ & $2.052$ & $2.043$ & $2.024$ & $2.045$ \\
$V_{\mathrm{s}}$ [km\,s$^{-1}$] & $5.1$ & $5.1$ & $5.4$ & $ - $\\
$Y_{\mathrm{s}}$ & 0.251 & 0.262 & 0.260 & 0.280 \\
$(Z/X)_{\mathrm{s}}$ & 0.0204 & 0.0205 & 0.0205 & 0.0205 \\
$[$Fe/H$]_{\mathrm{s}}$ &  $-0.05$ & $-0.05$ &  $-0.05$ & $-0.05$ \\
$\Delta \nu_{0}$ [$\mu$Hz] & $55.41$  & $55.59$ & $55.57$  & $55.56$\\
\hline
\end{tabular}
\end{center}
\end{table*}

\begin{figure}[htb!]
 \resizebox{\hsize}{!}{\includegraphics{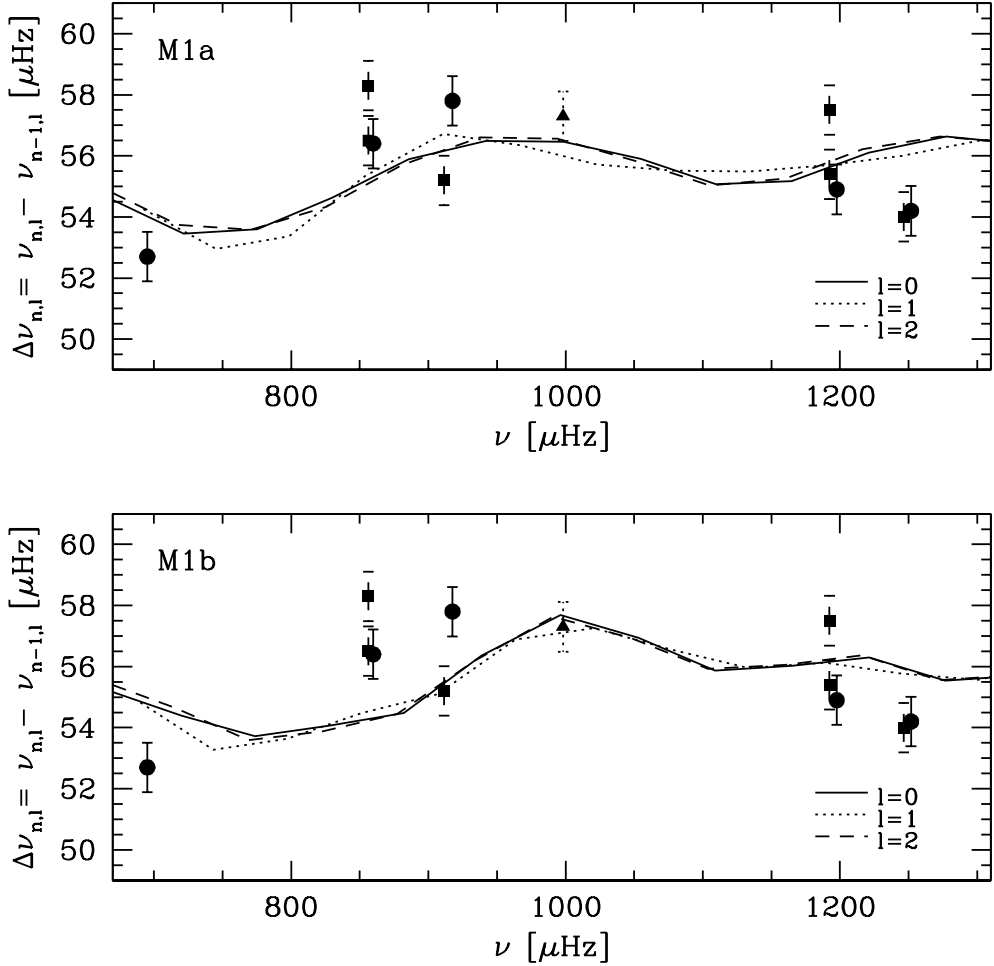}}
  \caption[]{Large spacing versus frequency for the models with a mass of 1.497~$M_{\odot}$. 
  The M1a model includes rotation and atomic diffusion, while the M1b model has been computed without rotation and diffusion
  (see Table~\ref{tab:res}).
  The dots indicate the observed values of the large spacing with an uncertainty on
  individual frequencies estimated to half the time resolution.}
  \label{gd_M1}
\end{figure}

\begin{figure}[htb!]
 \resizebox{\hsize}{!}{\includegraphics{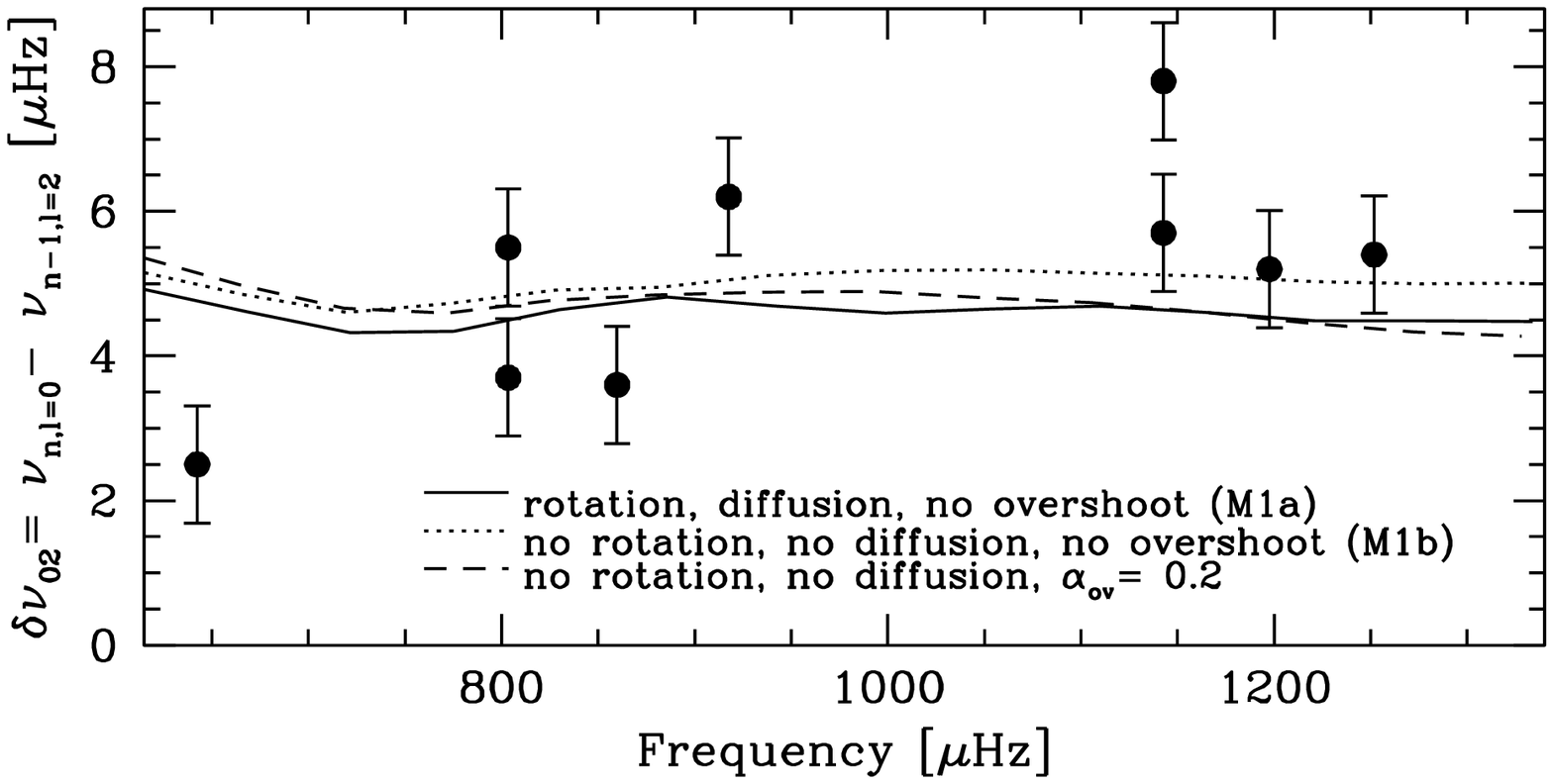}}
  \caption[]{Small spacing $\delta \nu_{02}$ versus frequency for the three models with a mass of 1.497~$M_{\odot}$.  
  The continuous line corresponds to the model including rotation and atomic diffusion but no overshooting
  (model M1a). The dotted line denotes the model without rotation, atomic diffusion nor overshooting (model M1b), while 
  the dashed line corresponds to a model computed without rotation or atomic diffusion, but with overshooting 
  ($\alpha_{\mathrm{ov}}=0.2$). 
  The dots indicate the observed values with an uncertainty on individual frequencies estimated to half the time resolution.}
  \label{pte}
\end{figure}

Concerning the asteroseismic features of this model, the theoretical variation of the large spacings $\Delta \nu_{\ell}$ for $\ell=0,1,2$ with 
frequency was compared to the observations (Fig.~\ref{gd_M1}). Table~\ref{tab:res} and Fig.~\ref{gd_M1} show that the mean large spacing of the M1a model
is in good agreement with the observed value of $55.5 \pm 0.5$\,$\mu$Hz. Moreover, 
one can see that the M1a model reproduces the observed variation of $\Delta \nu_{\ell}$ with frequency. 
However, the dispersion of the observed large spacings
around the theoretical curves is slightly greater than expected taking an uncertainty of 0.57\,$\mu$Hz, half of the time resolution,
on the frequency determination.
The comparison of the theoretical and observed values of the small spacing $\delta \nu_{02} \equiv \nu_{n+1,\ell=0}- \nu_{n,\ell=2}$ 
between $\ell=0$ and $\ell=2$ modes is given in Fig.~\ref{pte}. 
The mean small spacing of the M1a model and the theoretical variation of this spacing with frequency are compatible with the observed values.
However, Fig.~\ref{pte}
clearly shows that the observed small spacings are unfortunately not accurate enough to provide strong constraints to stellar models.
This is especially true for the couple of points near 799 and 1136\,$\mu$Hz that are either split by rotation or the secondary peaks are due to noise (see ECBB04).  
Finally, we compare the theoretical $p$--mode frequencies of the M1a model to the observed ones by plotting the echelle diagram (Fig.~\ref{ech}).
In this figure, the systematic difference $\langle D_{\nu}\rangle$ \citeaffixed{eg04a}{see Sect.~3.2.2 of}
between theoretical and observed frequencies has been taken into account. Indeed,
a linear shift of a few $\mu$Hz between theoretical and observational frequencies is perfectly acceptable, due to the fact that the exact values
of the frequencies depend on the details of the star's atmosphere where the pulsation is non--adiabatic.

\begin{figure}[htb!]
 \resizebox{\hsize}{!}{\includegraphics{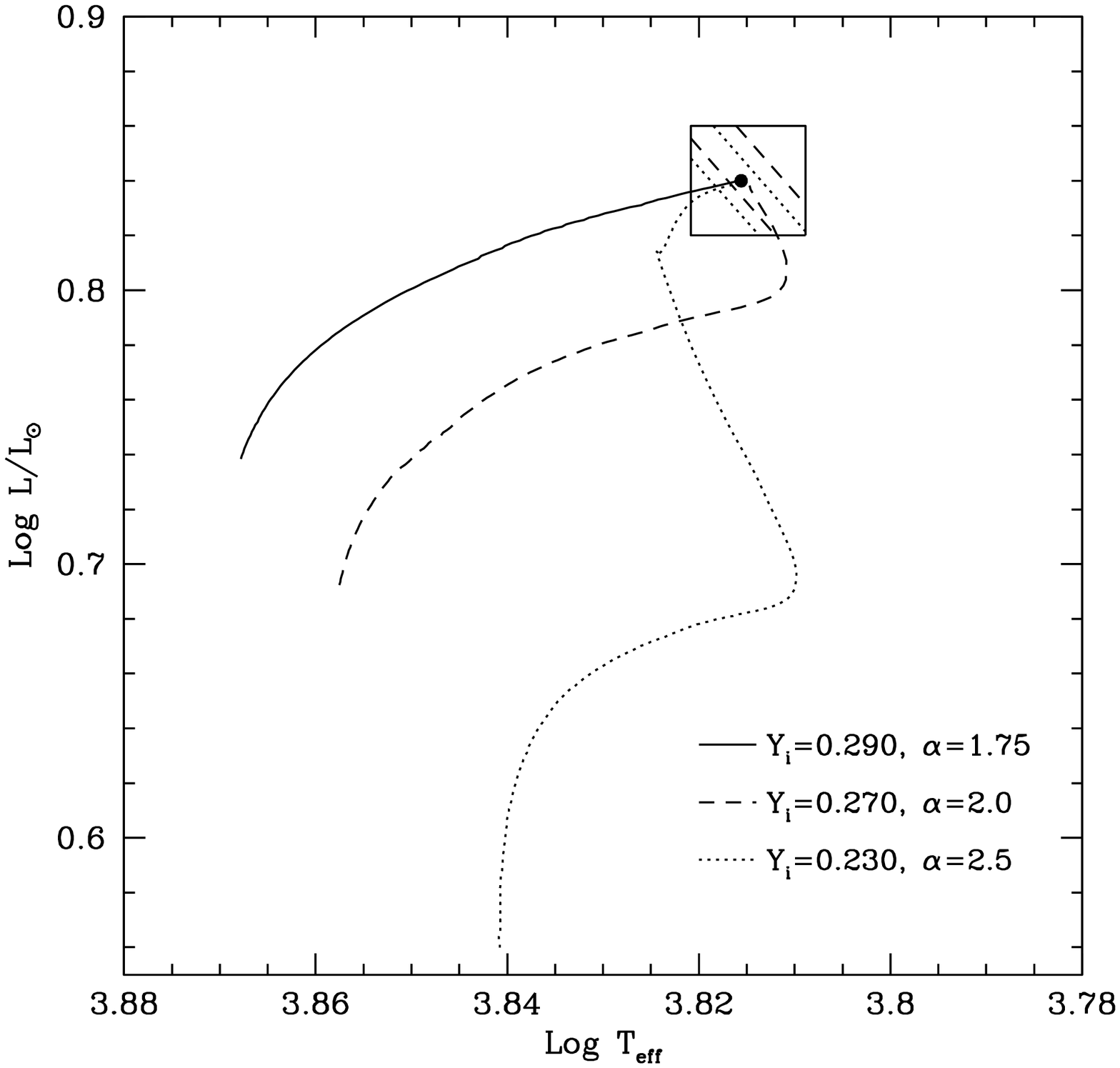}}
  \caption[]{Evolutionary tracks in the HR diagram for three models of 1.497~$M_{\odot}$ with different values of the initial helium abundance $Y_\mathrm{i}$
  and the convection parameter $\alpha$. 
  All models include atomic diffusion and rotation.
  This shows that a decrease in the initial helium abundance $Y_\mathrm{i}$ can be compensated by an increase of the convection parameter $\alpha$
  in order to reach the same location in the HR diagram.}
  \label{dhr_deg}
\end{figure}

\begin{figure}[htb!]
 \resizebox{\hsize}{!}{\includegraphics{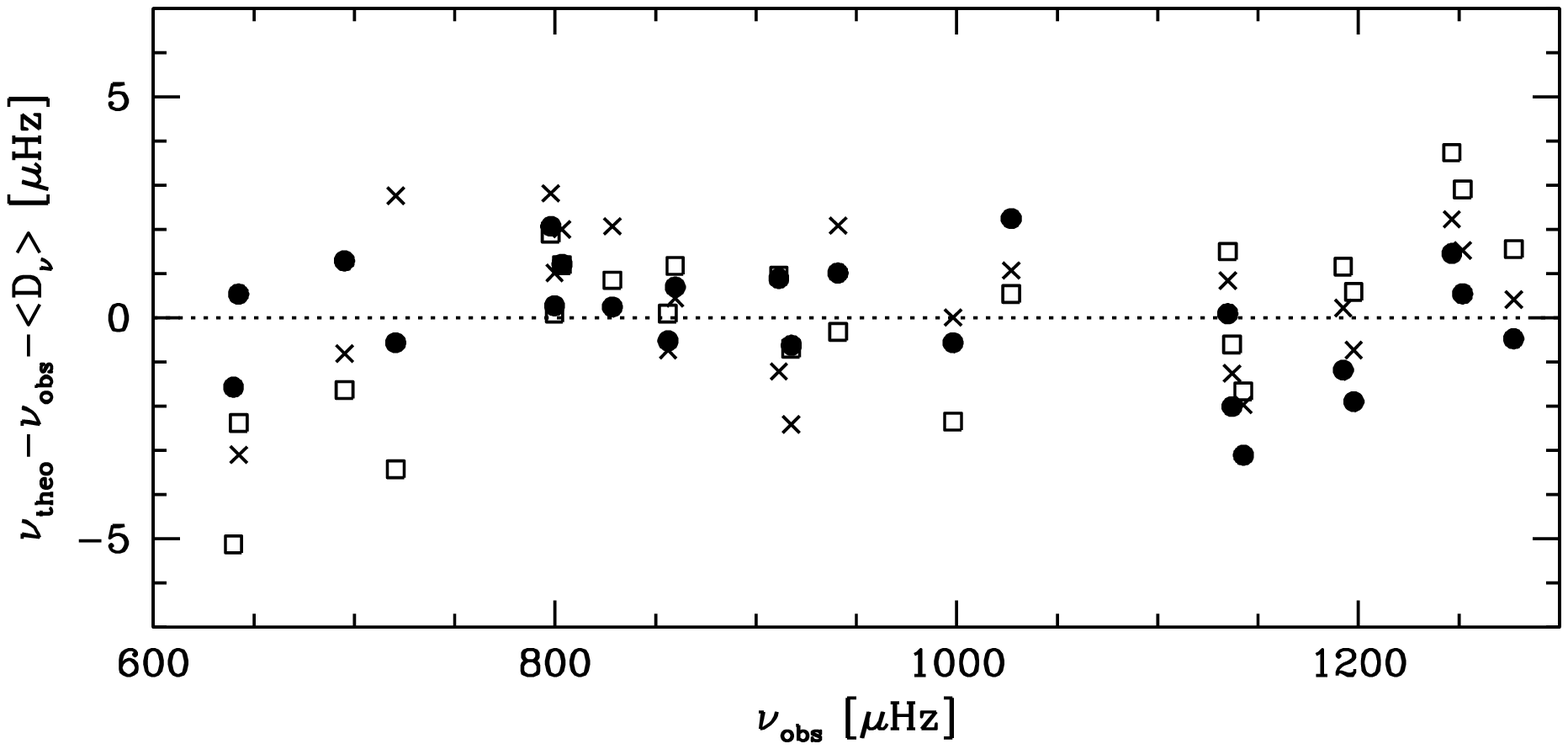}}
  \caption[]{Differences between calculated and observed frequencies for the models with 1.497\,$M_{\odot}$ including rotation and atomic diffusion
 with different values of the initial helium abundance $Y_\mathrm{i}$ and the convection parameter $\alpha$.
 Filled circles correspond to the model with $Y_\mathrm{i}=0.290$ and $\alpha=1.75$ (model M1a), while open squares and crosses designate
 models with $Y_\mathrm{i}=0.270$, $\alpha=2.0$ and $Y_\mathrm{i}=0.230$, $\alpha=2.5$, respectively. 
  The systematic shifts $\langle D_{\nu}\rangle$ have been taken into account
  (see text for more details).}
  \label{theobs}
\end{figure}

Model M1a corresponds to the best solution which minimizes the $\chi^2$ functional of Eq.~\ref{eq1}.
However, the minimization shows that many models with different values of the initial helium abundance and the
mixing--length parameter $\alpha$ have a $\chi^2$ only slightly larger than the one of the
M1a model, and constitute therefore also good models of \astrobj{Procyon~A}. This is due to the fact that a decrease/increase of the 
initial helium abundance $Y_{\mathrm{i}}$ can be compensated by an increase/decrease of the mixing--length parameter $\alpha$
to match the observed position of \astrobj{Procyon~A} in the HR diagram (see Fig.~\ref{dhr_deg}).
Thus, we obtain a series of models with approximately the same non--asteroseismic features as those of the M1a model.  
Moreover, the mean large spacing of these models (which is the only asteroseismic quantity included in our $\chi^2$ functional)
is also very close to the value of the M1a model, since it mainly depends on the star's mean
density and hence on its radius given that the models have the same mass of 1.497\,$M_{\odot}$. 
This explains why there is a series of models which well reproduced the global stellar parameters considered in our
minimization. Amongst these models, the M1a model is in slightly better agreement with these observables.
Moreover, when individual theoretical and observational asteroseismic frequencies are compared, one finds that the M1a model
is also in slightly better agreement with the asteroseismic observations than the other models (see Fig.~\ref{theobs}).
However, given the limited accuracy of the observations, and especially of the observed small spacings (which are needed
to differentiate models with the same position in the HR diagram but different ages), these differences cannot be considered
as really significant.     

To investigate the effects of rotation on the structure of the models and therefore on their $p$--mode frequencies, 
we decided to redo the entire calibration without including the effects of rotation and atomic diffusion. 
By performing the same minimization on this grid of standard models, we found the solution
$t=1.77 \pm 0.30$\,Gyr, $\alpha=1.60 \pm 0.35$, 
$Y_{\mathrm{i}}=0.280 \pm 0.010$ and $(Z/X)_{\mathrm{i}}=0.0205 \pm 0.0015$. The position and the evolutionary track of this model 
(denoted model M1b in the following) in the HR diagram are shown in Fig.~\ref{dhr} and \ref{dhr_ov}. 
The characteristics of this model are reported in Table~\ref{tab:res}.
First, we note that the global parameters of this M1b model are very similar to the ones of the M1a model: same effective temperature and
approximately same luminosity, radius and mean large spacing as those of the M1a model. Moreover, the ages of both models are very similar.
Of course, the initial chemical composition of the models is different, since the M1b model does not take into account the rotationally induced
chemical mixing and atomic diffusion.
Moreover, the mixing--length parameter of both models is approximately equal to the solar calibrated value determined by using the same input physics
($\alpha_{\odot}=1.75$ with rotation and atomic diffusion, and $\alpha_{\odot}=1.59$ without rotation and diffusion).  

Concerning the asteroseismic features of the M1b model, 
the comparison between observed and theoretical variation of the large spacings $\Delta \nu_{\ell}$ is shown on Fig.~\ref{gd_M1}. 
This figure and Table~\ref{tab:res} show that the mean large spacing and the variation of the large spacings with frequency of the M1b model
are in good agreement with the observations. Thus, 
one can see that there are no significant differences between the large spacings of both M1 models; the mean large spacing of the M1b model
is only slightly larger than the one of the M1a model, which results from the smaller radius of this model compared to the one of the M1a model.
 
To investigate the effects of rotation on the $p$--mode frequencies,
it is much more interesting to consider the small spacings $\delta \nu_{02} \equiv \nu_{n+1,\ell=0}- \nu_{n,\ell=2}$ 
which are mainly sensitive to the stellar core. 
Indeed, as already pointed out by \citeasnoun{di01} and \citeasnoun{pr02}, these spacings may enable a test for overshoot
if sufficiently precise frequencies could be determined. 
To check this, we also computed a model of \astrobj{Procyon~A} without rotation and atomic diffusion, 
but with an overshoot of the convective core into the radiative zone on a distance 
$d_{\mathrm{ov}} \equiv \alpha_{\mathrm{ov}} \min[H_p,r_{\mathrm{core}}]$. 
Following \citeasnoun{sc92}, we fixed the value of $\alpha_{\mathrm{ov}}$ to 0.2,
which is the amount of overshooting usually chosen for non--rotating stellar
models. The evolutionary track of this model is shown on Fig.~\ref{dhr_ov}. 
We find that the values of the small spacings of stellar models including overshooting are slightly smaller
than those computed for models without overshooting. 
This can be seen in Fig.~\ref{pte}, where the small spacings of two non--rotating models (one with and the other without overshooting)
with the same averaged large spacing are compared. 
In the same way, it is interesting to note that these differences also exist between the non--rotating model without overshooting (model M1b)
and the M1a model calculated with rotation and atomic diffusion, but without overshooting.
Indeed, Fig. \ref{pte} shows that the non--rotating model (M1b) exhibits slightly higher  values of the small spacings than the rotating one (M1a).
Thus, we see that $p$--mode frequencies indicate that the inclusion of rotation mimics an increase of the overshoot parameter. 
This results from the fact that, for sufficiently low initial velocities ($v_{\mathrm{ini}} \lesssim 300$ km\,s$^{-1}$) 
rotation acts as an overshoot, extending the main sequence tracks towards lower effective temperatures (see Fig. \ref{dhr_ov}).

Even if these theoretical considerations are interesting to point out, it is evidently not necessary to say that, from an observational point of view, 
we are very far from
the precision required to discuss these subtle differences. 
As mentioned above, we can only conclude that the observed frequency differences between $\ell=0$ and $\ell=2$ modes are compatible with theoretical
expectations, but are not accurate enough to enable us to test the input physics of our stellar models.

\subsection{Models with a mass of $1.465\,M_{\odot}$}

After having calculated stellar models with the mass of 1.497\,$M_{\odot}$ determined by \citeasnoun{gi00}, 
we computed a grid of models with a mass of $1.465\,M_{\odot}$. This mass was deduced using only the WFPC2 measurements with the
parallax of \citeasnoun{gi00}. Accordingly, the value of the observed linear radius remains unchanged ($2.067 \pm 0.028$\,$R_{\odot}$).
All models include shellular rotation and atomic diffusion.

Using this new value for the mass, we found the following solution:
$t=1.89 \pm 0.30$\,Gyr, $\alpha=1.90 \pm 0.30$, $V_{\mathrm{i}}=17 \pm 8$\,km\,s$^{-1}$, 
$Y_{\mathrm{i}}=0.295 \pm 0.015$ and $(Z/X)_{\mathrm{i}}=0.0229 \pm 0.0015$. The position of this model 
(denoted model M2 hereafter) in the HR diagram is shown in Fig.~\ref{dhr}, while 
its characteristics are given in Table~\ref{tab:res}.

\begin{figure}[htb!]
 \resizebox{\hsize}{!}{\includegraphics{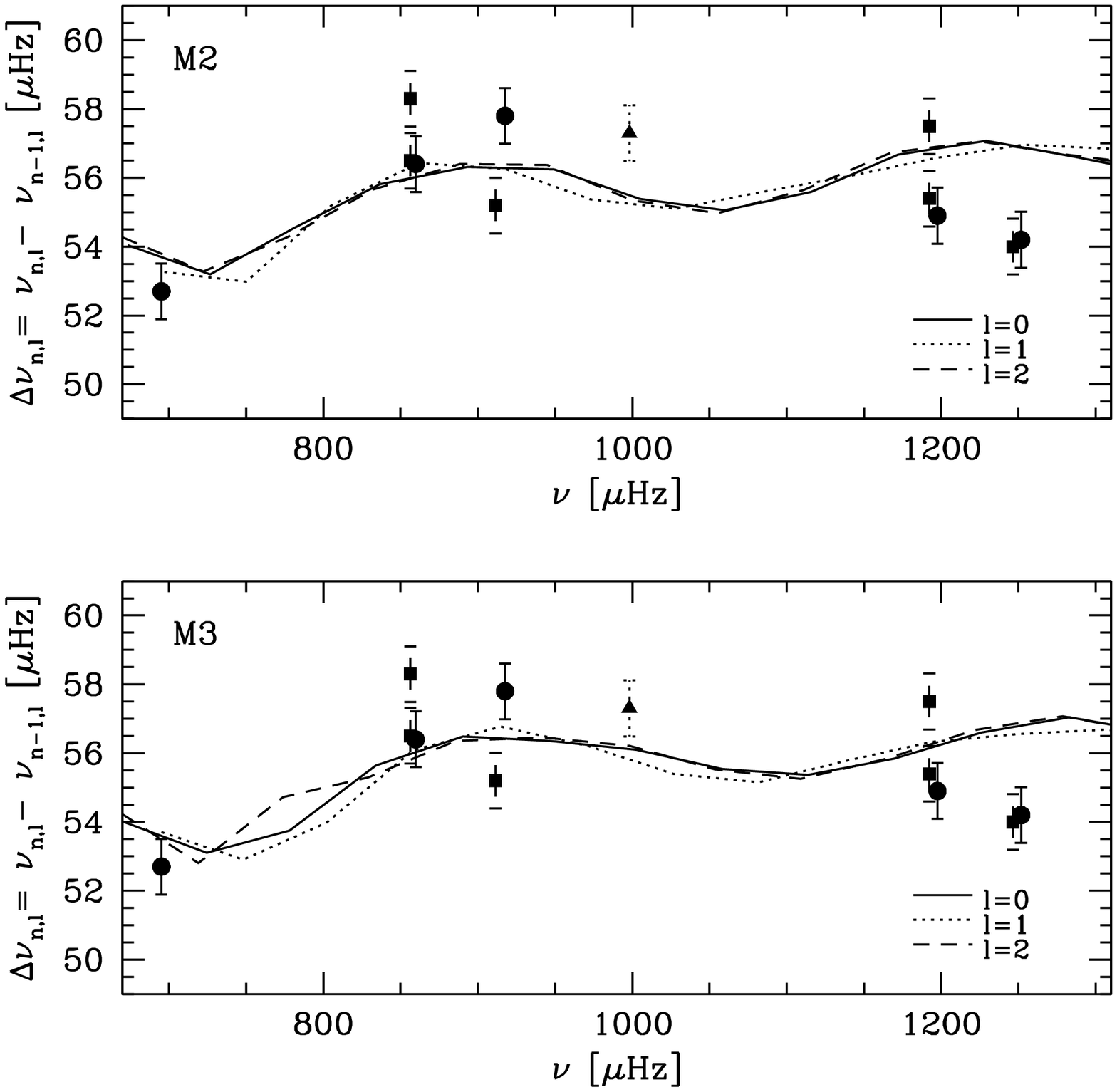}}
  \caption[]{Large spacing versus frequency for the models with masses
  of 1.465~$M_{\odot}$ (model M2) and 1.423~$M_{\odot}$ (model M3). 
  Both models include the effects of rotation and atomic diffusion.
  The dots indicate the observed values of the large spacing with an uncertainty on
  individual frequencies estimated to half the time resolution.}
  \label{gd_M23}
\end{figure}

We note that the modeling parameters of this M2 model are very similar to the ones of the M1a model.
Indeed, the initial chemical composition of both models are approximately identical, with a nearly solar initial
metallicity which decreases during the star evolution to reach the observed value of $[$Fe/H$]_{\mathrm{s}}=-0.05$.
The mixing--length parameter and the initial velocity are only slightly larger for the M2 model than for the M1a. 
Due to the lower mass, the age of the M2 model is somewhat larger than the one of the M1 models.
Note that the large error of 8\,km\,s$^{-1}$ on the initial velocity is due to the large error on the observed
surface rotational velocity (relative error of about 35\,\%), while the large error on the mixing--length parameter $\alpha$
simply reflects the fact that the models are not very sensitive to a change of this parameter.

\begin{figure}[htb!]
 \resizebox{\hsize}{!}{\includegraphics{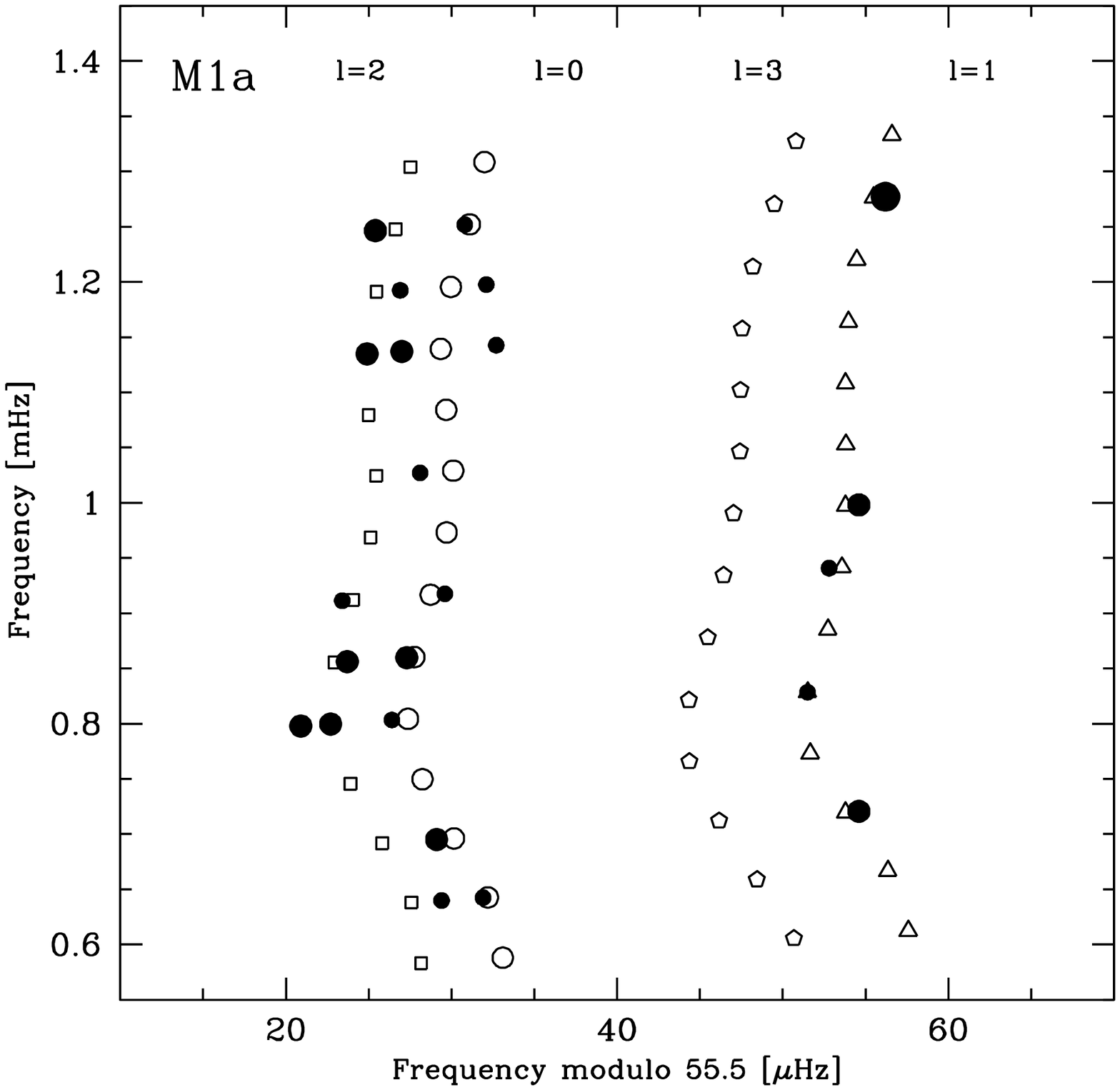} \includegraphics{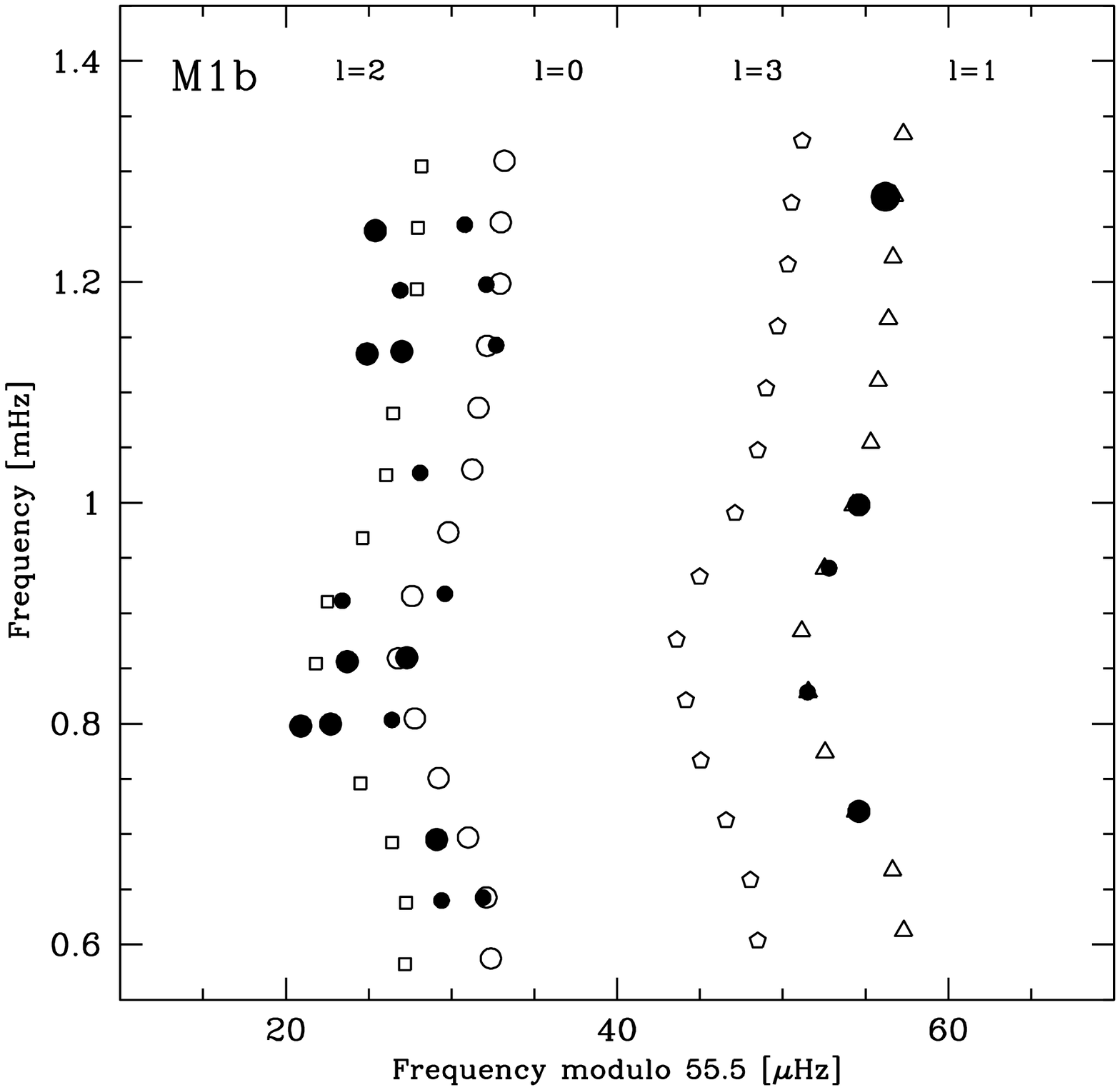}}
 \resizebox{\hsize}{!}{\includegraphics{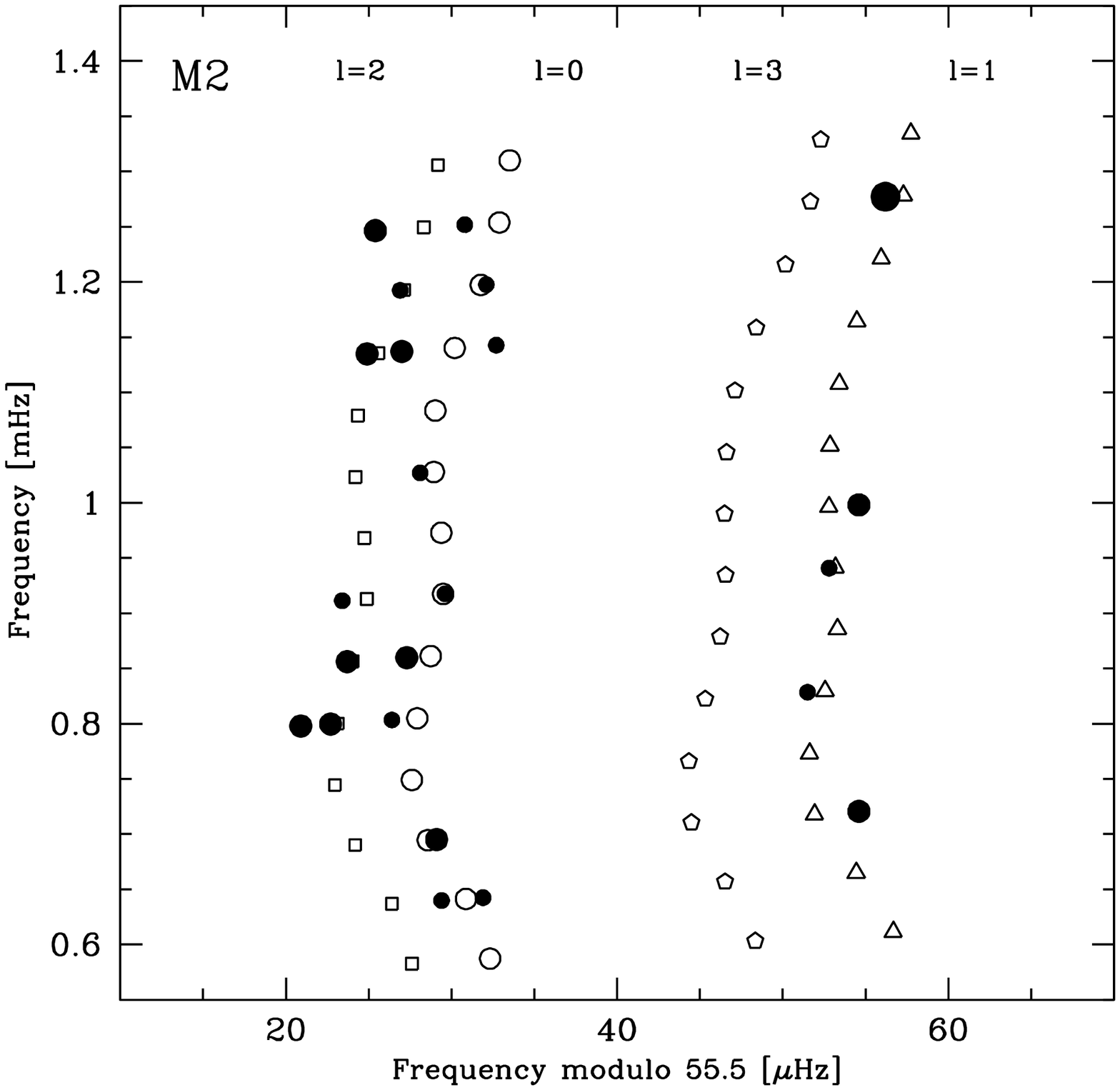} \includegraphics{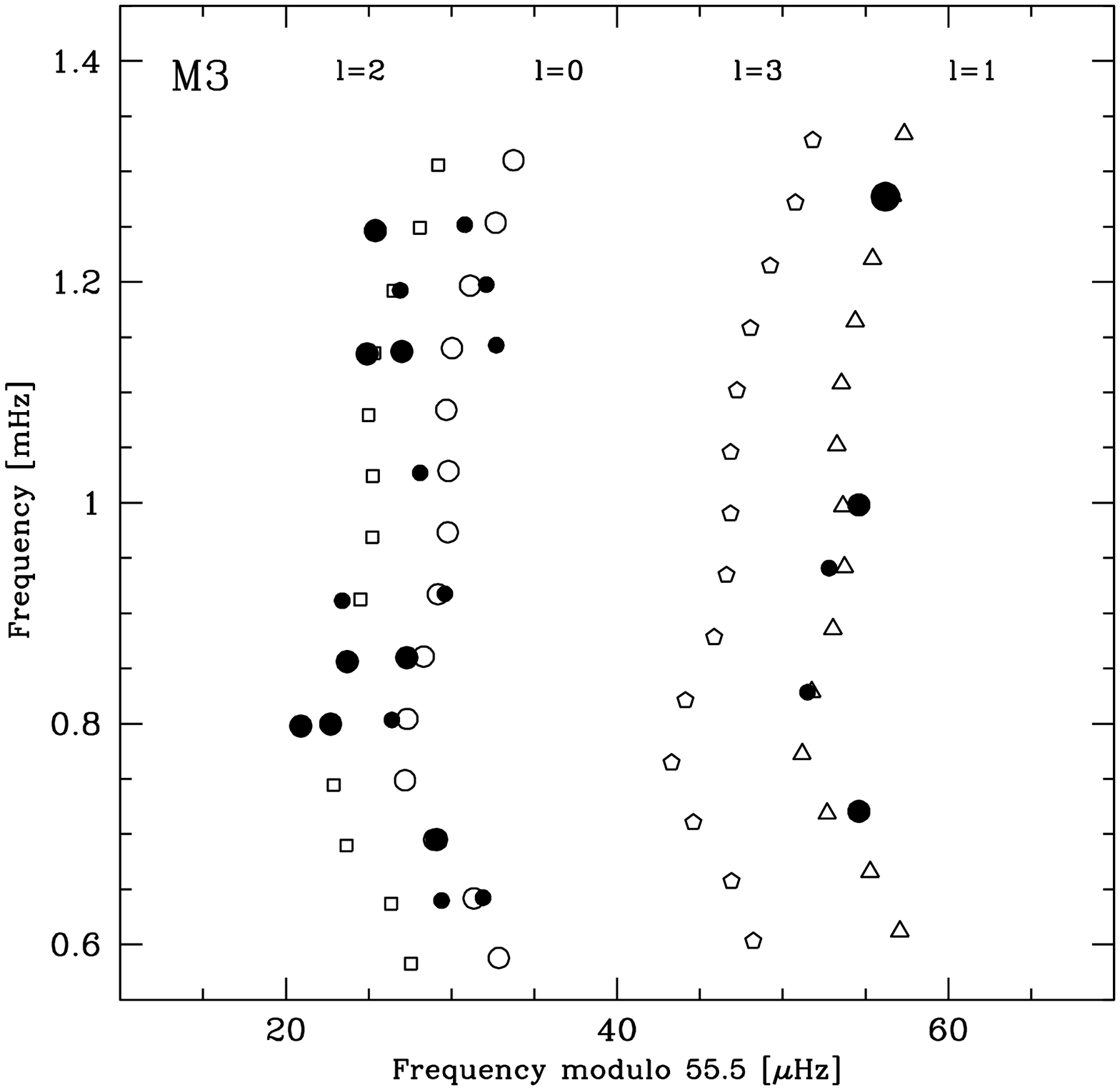}}
  \caption[]{Echelle diagrams for the four models of \astrobj{Procyon~A} listed in Table~\ref{tab:res}.
Open symbols refer to theoretical
frequencies, while the filled circles correspond to the observed frequencies (see ECBB04). 
Open circles are used for modes with $\ell=0$, triangles for $\ell=1$, squares for $\ell=2$ and pentagons for $\ell=3$.
The models with a mass of 1.497\,$M_{\odot}$ are in better accordance with the asteroseismic data than the less massive M2 and
M3 models.}
  \label{ech}
\end{figure}

Using this lower mass leads to a model whose global parameters still reproduce the observational
constraints. However, the agreement between observed and theoretical non--asteroseismic parameters
is slightly better for the more massive M1 models, than for this model. Concerning the asteroseismic 
features, one can see that the observed large spacings are well reproduced by the M2 model (see Fig.~\ref{gd_M23}).  
Finally, the echelle diagram of Fig.~\ref{ech} shows that the theoretical frequencies of the M2 model are compatible with the asteroseismic observations,
even if the agreement is slightly better for the more massive M1 models than for this model. 
However, the limited accuracy of the seismological measurements (especially for the observed small spacings $\delta \nu_{02}$)
does not enable us to formally reject the M2 model.

\subsection{Models with a mass of $1.423\,M_{\odot}$}

Finally, we computed a grid of stellar models including shellular rotation and atomic diffusion with a mass of $1.423\,M_{\odot}$.  
This mass was deduced by using only the WFPC2 measurements with the
Hipparcos parallax (see Sect.~\ref{astro}). Accordingly, the observed radius of \astrobj{Procyon~A} is decreased from 
$2.067 \pm 0.028$\,$R_{\odot}$ to $2.048 \pm 0.025$\,$R_{\odot}$.

By redoing the calibration with this new mass and new observational constraint on the radius,
we found the solution
$t=2.18 \pm 0.30$\,Gyr, $\alpha=1.80 \pm 0.30$, $V_{\mathrm{i}}=17 \pm 8$\,km\,s$^{-1}$, 
$Y_{\mathrm{i}}=0.295 \pm 0.020$ and $(Z/X)_{\mathrm{i}}=0.0231 \pm 0.0015$. The position of this model 
(denoted model M3 in the following) in the HR diagram is shown in Fig.~\ref{dhr}, while 
the characteristics of this model are given in Table~\ref{tab:res}.
The modeling parameters of this M3 model are also very similar to the ones of the other rotating models, 
apart from a larger age, which results directly from the lower mass of the star.

One can see from Table~\ref{tab:res} and Fig.~\ref{dhr} that the agreement between the theoretical and observed global parameters
of \astrobj{Procyon~A} is somewhat better for the more massive models (models M1 and M2) than for this M3 model.
This is especially true for the radius of $2.024$\,$R_{\odot}$ which lies in the lower
border of the observed box (see Fig.~\ref{dhr}). This is due to the fact that the radius of this model has to be lower
than the radii of the other models in order to compensate the mass decrease and to keep a value of the mean large spacing
close to the observed value of 55.5\,$\mu$Hz.
Besides, one can see on Fig.~\ref{gd_M1} that the theoretical large spacings of the M3 model are in good agreement with the observed ones.
The echelle diagram of Fig.~\ref{ech} shows that the $p$--modes frequencies of the M3 model are compatible with the asteroseismic observations,
even if the agreement is better for the more massive M1 models than for this model.

\section{Conclusion}

By combining all
non--asteroseismic observables with the asteroseimic measurements of
ECBB04, we find that the observed mean large spacing of $55.5 \pm 0.5$\,$\mu$Hz 
favours a mass of 1.497\,$M_{\odot}$ for \astrobj{Procyon~A}.
Indeed, the lower masses of 1.465 and 1.423\,$M_{\odot}$ lead to models which are in 
less good agreement with the observational constraints.
We also determine the following global parameters of \astrobj{Procyon~A}:
an age of $t=1.72 \pm 0.30$\,Gyr, an initial helium mass fraction $Y_{\mathrm{i}}=0.290 \pm 0.010$, 
a nearly solar initial metallicity $(Z/X)_{\mathrm{i}}=0.0234 \pm 0.0015$
and a mixing--length parameter $\alpha=1.75 \pm 0.40$.

Note that the lower value of $53.6 \pm 0.5$\,$\mu$Hz reported by \citeasnoun{ma04}
favours the lower mass of 1.423\,$M_{\odot}$ \cite{ke04}.
This can be immediately understood by recalling that the mean large spacing is proportional to the square root of the star's mean density.
Since the radius of \astrobj{Procyon~A} is now precisely determined (thanks to interferometric measurements),
a lower value of the mean large spacing results in a lower value of the mass.  

We also show that the effects of rotation on the inner structure of the star may be revealed by asteroseismic observations,
if precise measurements of the small spacings between $\ell=2$ and $\ell=0$ modes can be determined. Indeed,
there are subtle differences in the frequency dependence of the small spacing between rotating and non--rotating models, which results
from changes in the size and in the outer border of the star's convective core. As expected for small initial velocities, these rotational effects
are found to mimic the effects due to an overshoot of the convective core into the radiative zone.   
Unfortunately, the asteroseismic observations now available for \astrobj{Procyon~A} are not accurate enough to enable us to discuss these
subtle differences. 

We conclude that the combination of all non--asteroseismic measurements with 
existing asteroseismic observations puts important constraints on the global parameters of \astrobj{Procyon~A}, but that  
more accurate asteroseismic data are needed to really
test the physics of the stellar models.

\section*{Acknowledgements}
We would like to thank J. Christensen--Dalsgaard for providing us with the Aarhus adiabatic pulsation code.
We also thank A.~Maeder, G.~Meynet, C.~Charbonnel and S.~Talon for helpful advices.
This work was partly supported by the Swiss National Science Foundation.

\end{document}